\documentstyle[pra,aps,floats]{revtex}
\newcommand{\be}{\begin{equation}}
\newcommand{\ee}{\end{equation}}  
\newcommand{\ba}{\begin{eqnarray}}
\newcommand{\ea}{\end{eqnarray}}

         \newcommand{\qb}{\begin{equation}}
         \newcommand{\qba}{\begin{eqnarray}}
         \newcommand{\qbas}{\begin{eqnarray*}}
         \newcommand{\qe}{\end{equation}}
         \newcommand{\qea}{\end{eqnarray}}
         \newcommand{\qeas}{\end{eqnarray*}}
         
\input epsf

\begin{document}
\twocolumn[\hsize\textwidth\columnwidth\hsize\csname
@twocolumnfalse\endcsname

\title{Quantum origin of classical properties within the modal interpretations}
\author{Gyula Bene\cite{email1}}
\address{Institute of Theoretical Physics, E\"otv\"os University,\\
P\'azm\'any P\'eter s\'et\'any 1/A, H-1117 Budapest, Hungary\footnote{z3}}
\date{\today}
\maketitle

\begin{abstract}%
An example is presented when
decoherence and quantum interference gives rise to narrow eigenstates
(in coordinate representation)
for the reduced density matrix of macroscopic quantum systems.
On the basis of modal interpretations this means
the emergence of classical properties. 
   
\end{abstract}

\vskip2pc] % end of twocolumn

\narrowtext

Modal interpretations aim at extracting a consistent pysical
picture out of the formalism of quantum mechanics rather than relying on 
assumptions about an {\em a priori} classical world\cite{VanFraassen},
\cite{Dieks},\cite{Vermaas_Dieks}. 
The Dieks-Vermaas version of the modal interpretations utilizes
the Schmidt states, i.e., the eigenstates of the reduced density matrix,
and identifies them with the actually existing, physical states. 
The significance of the Schmidt bases has previously been emphasized 
- in connection with decoherence and Everett's many worlds interpretation 
- by Zeh\cite{Zeh}. The Schmidt basis plays a decisive role also
in the relational
version of the modal interpretation\cite{Bene97} which incorporates 
the idea of the relativity of the states\cite{Everett}, \cite{Mermin}, \cite{Rovelli}.
A serious objection raised against modal interpretations has been 
the observation
that in case of some simple systems like a free point mass or an oscillator
the eigenstates of the reduced density matrix do not become narrow, 
although the reduced density matrix becomes almost diagonal
in coordinate space 
due to decoherence\cite{Bacciagaluppi}. 
Here I present an example where the eigenstates do become narrow in coordinate
space, suggesting that the previous objections are based on anomalous
special situations.

The model used is a massive particle moving along 
an interval (i.e., between infinite potential walls) 
in one dimension. This plays the role of a macroscopic system.
The initial condition is a Gaussian wave packet. 
Initially the paticle moves freely, afterwards it interacts
with the environment which leads to decoherence making the reduced
density matrix almost diagonal in coordinate space, without changing
the diagonal elements. This system has been investigated numerically
in case of the free motion, while the subsequent decoherence has been 
taen into account with the introduction of a decoherence factor
$\exp\left(-(x-x')^2/d^2\right)$. The decoherence length is assumed 
to be much smaller than the deBroglie wavelength of the ``macroscopic''
particle. The remarkable fact is displayed in the figure, namely,
that interference leads to strong spatial oscillations at the length
scale of the deBroglie wavelength. The (almost) zero minima imply
that after decoherence the reduced density matrix
\ba
\rho(x,x')=\psi(x)\psi^*(x')\exp\left(-(x-x')^2/d^2\right)
\ea    
has a block diagonal structure, so that each block has a size
comparable with the deBroglie wavelength. Consequently, the eigenstates
of the reduced density matrix will have the same widths, too. Note that the 
interference structure changes in time, however, qualitatively it behaves
similarly, i.e., the oscillations and the nearly zero minima always occur
after the initial wave packet has spreaded over the whole interval.
The time evolution has been solved numerically via fast Fourier transform.
The correctness and precision has been carefully checked, among others,
by changing the sign of the time to get back the initial state. 
The parameters corresponding to the figure are $L=1$ (length of the interval),
$\hbar=1$ (Planck's constant), $p_0=30$ (initial momentum), 
$q_0=0.5$ (initial position),
$\sigma=0.05$ (initial width of the wave function), $m=1$ (mass),
$t=0.5$ (time). 

\begin{figure}
\begin{center}
\noindent\epsfbox{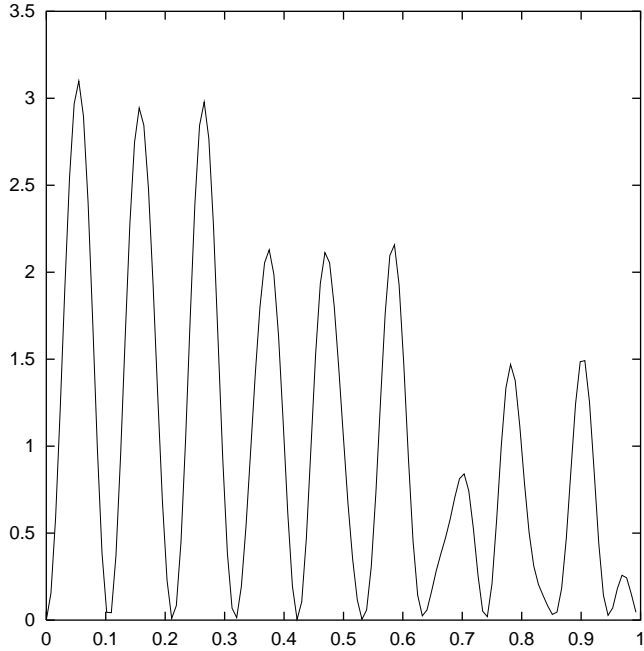}
\end{center}
\caption{
Typical interference pattern in the modulus 
square of the particle's wave function in coordinate space. The initial Gaussian wave packet
spreaded over the whole interval. 
}
\end{figure} 

In this case classical behavior is due to an interplay between
quantum interference and decoherence. The most important ingredients 
have been
the zeros of the wave function and the strong decoherence
which makes the reduced density matrix much narrower in coordinate space
than the distance between the zeros.
 
Our preliminary results show that the same effect occurs in two
dimensions as well.

{\bf Acknowledgements}

Several enlightening discussions with  
Andreas Bringer, Dennis Dieks, Gert Eilenberger, Michael Eisele, G\'eza Gy\"orgyi,
Frigyes K\'arolyh\'azy, Hans Lustfeld, Roland Omnes, Zolt\'an Perj\'es, 
L\'aszl\'o Szab\'o, Mikl\'os R\'edei and
 Pieter Vermaas are gratefully acknowledged.

This work has been partially supported by the Hungarian Aca\-demy of
 Sciences
 under Grant No. OTKA T 029752, T 031 724  and the J\'anos Bolyai Research Fellowship.

\end{document}